\documentclass[12pt]{article}
\def\psl{\hbox{/\kern-.5800em$p$}}

\def\gappeq{\mathrel{\rlap {\raise.5ex\hbox{$>$}}
{\lower.5ex\hbox{$\sim$}}}}
\def\lappeq{\mathrel{\rlap{\raise.5ex\hbox{$<$}}
{\lower.5ex\hbox{$\sim$}}}}
\begin{document}
\pagestyle{empty}
\begin{flushright}
UMN-TH-2227/03\\
December 2003
\end{flushright}
\vspace*{5mm}

\begin{center}
{\Large\bf Dirac Neutrino Masses with\\}
\vspace{0.2cm}
{\Large\bf Planck Scale Lepton Number Violation}
\vspace{1.0cm}

{\sc Tony Gherghetta}\\
\vspace{.5cm}
{\it\small {School of Physics and Astronomy\\
University of Minnesota\\
Minneapolis, MN 55455, USA}}\\
\end{center}

\vspace{1cm}
\begin{abstract}

It is shown how pure Dirac neutrino masses can naturally occur
at low energies even in the presence of Planck scale lepton number 
violation. The geometrical picture in five dimensions assumes that
the lepton number symmetry is explicitly broken on the Planck 
brane while the right-handed neutrino is localised on the TeV brane. This 
physical separation in the bulk causes the global lepton number 
to be preserved at low energies. A small wavefunction overlap between the 
left-handed and right-handed neutrinos then naturally leads to a small Dirac
Yukawa coupling. By the AdS/CFT correspondence there exists a purely 
four-dimensional dual description in which the right-handed neutrino 
is a composite CFT bound state. The global lepton number is violated
at the Planck scale in a fundamental sector whose mixing into the composite
sector is highly suppressed by CFT operators with large anomalous dimensions.
A similar small mixing is then also responsible for
generating a naturally small Dirac Yukawa coupling between the fundamental 
left-handed neutrino and the composite right-handed neutrino.

\end{abstract}

\vfill
\begin{flushleft}
\end{flushleft}
\eject
\pagestyle{empty}
\setcounter{page}{1}
\setcounter{footnote}{0}
\pagestyle{plain}

%============================================================

\section{Introduction}

Recent neutrino oscillation experiments have spectacularly confirmed 
that neutrinos have a small nonzero mass. However it remains
an open question as to whether the neutrino mass is of the Majorana or 
Dirac type. Historically the preference is that they are of the Majorana 
type, because a small mass can be elegantly explained by the see-saw 
mechanism~\cite{seesaw}. But this comes at the expense of introducing a new 
mass scale below the Planck scale, which presumably is related to the 
usual GUT scale. Since there is no experimental evidence for this picture 
it behooves us to consider the possibility that neutrinos could be Dirac 
particles.

The usual argument against Dirac neutrino masses consists
of requiring unnaturally small Yukawa couplings, although this argument
seems presumptuous especially given the fact that we do not 
understand why the electron Yukawa coupling is so small. Furthermore, one 
must impose a global lepton number symmetry to forbid a Majorana mass term 
for the right-handed neutrino. However this global symmetry can be badly 
broken by black hole physics, and consequently for any
consistent theory of quantum gravity. In fact, in string theory it is not
possible to obtain an additive global conservation law~\cite{witten}. 
These symmetry violating effects at an ultraviolet 
(UV) scale $M$ can be parametrised 
by including all lepton number violating operators in the Lagrangian. 
If we introduce the right-handed Weyl neutrino $\nu_R$, which is a standard 
model singlet carrying one unit of lepton number, then the dimension three 
Majorana mass term
\begin{equation}
                   M (\nu_{R}\nu_R + c.c.)~,
\label{majterm}
\end{equation}
is the most relevant lepton number violating operator. At low energies
we can integrate out this right-handed neutrino state. This then leads 
to the usual seesaw mechanism and to the Majorana nature of the 
physical neutrino mass. Thus, in the presence of Planck scale 
lepton number violating effects it would seem to be 
difficult to maintain the Dirac nature of the neutrino.

In this paper we show that the above incompatibility can be naturally 
avoided in a warped extra dimension, or strongly coupled conformal
field theory (CFT). We will argue that if the UV physics 
violates the lepton number symmetry maximally, then this symmetry 
will still be restored in the infrared (IR), where the physical neutrino is 
a Dirac particle. So the usual argument that the Standard Model has an 
accidental global lepton number symmetry at low energies 
because all lepton number violating operators are irrelevant will be extended 
to the case of the Standard Model plus a right-handed neutrino. While it 
will be easy to visualise our scenario in a warped extra 
dimension, remarkably we will see that because of the AdS/CFT 
correspondence~\cite{adscft} the geometrical mechanism of the extra 
dimension can be completely interpreted in four dimensions where 
the right-handed neutrino is a CFT bound state\footnote{In purely four 
dimensions composite right-handed neutrinos generating naturally small 
Dirac masses were also considered in Ref.~\cite{ahg}.}. 
Since the CFT is a large 
$N$ strongly coupled gauge theory, the coupling of the composite state 
to the UV violation of lepton number involves a form factor $F(q^2)$ which
vanishes in the limit $q^2\rightarrow\infty$. In fact the operator
responsible for creating a right-handed neutrino acquires a large anomalous 
dimension. This causes the usual dimension three operator (\ref{majterm}) 
to be highly suppressed in the IR and the right-handed neutrino does not 
decouple. In other words lepton number violation is only characterised by 
irrelevant operators, leading to a theory containing a right-handed neutrino
with an accidental global lepton number symmetry in the IR.
A similar mechanism occurs for global supersymmetry, which can be maximally
broken on the Planck brane, while fields confined to the TeV brane (such 
as the Higgs scalar field) remain supersymmetric~\cite{gp4}. Other 
interesting possibilities have also been discussed in Refs~\cite{luty1,cnp,
luty2,strassler}. 

Neutrino masses in extra dimensions were first considered in flat space 
in Ref.~\cite{flat}. The generalisation to warped extra dimensions was 
studied by Grossman and Neubert~\cite{gn}, where a Majorana mass for 
the right-handed neutrino was forbidden by imposing a global lepton number 
symmetry. But since the right-handed neutrino is localised on the Planck brane 
one would expect that the global symmetry breaking effects at 
the Planck scale will lead to Majorana neutrino masses, and not Dirac masses.
More recently in Ref.~\cite{yn,hs}, a Majorana mass for the right-handed 
neutrino was introduced on the Planck brane in order to obtain Majorana 
neutrino masses of the right magnitude via the usual seesaw mechanism. 
Instead by localising the right-handed neutrino towards the TeV brane we 
will see that Dirac neutrino masses can be naturally generated even in the 
presence of a large lepton number violating term on the Planck brane. 
The effective right-handed Majorana mass can be naturally made to vanish.

Let us first begin with the geometrical picture of our setup in
five dimensions. This will make it easy to visualise how the Dirac
nature of the neutrino can be naturally obtained.
Consider a slice of AdS$_5$ where the fifth dimension is compactified on
an orbifold $S^1/Z_2$ of radius $R$ with $0 \leq y\leq \pi R$.
The metric solution is given by
\begin{equation}
   ds^2 = e^{-2 k |y|}\eta_{\mu\nu} dx^\mu dx^\nu + dy^2~,
\end{equation}
where $k$ is the AdS curvature scale, and the Minkowski metric 
$\eta_{\mu\nu}$ has signature $(-+++)$.
At the orbifold fixed points $y^\ast=0$ and $y^\ast=\pi R$ there 
are two 3-branes, the Planck brane and TeV brane, respectively. 
For each neutrino Weyl fermion, $\nu_{L,R}$ we will associate
a five dimensional (5D) Dirac fermion field $\Psi_{L,R}$ with components
\begin{equation}
    \Psi_i(x^\mu,y) = \left(\begin{array}{c} \psi_{1\,i}(x^\mu,y)\\
                               \bar\psi_{2\,i}(x^\mu,y)\end{array}\right)~,
\end{equation}
where $i=L,R$. We will assume a separation of variables for the wavefunctions
$\psi_{k\,i}(x,y) = 1/\sqrt{\pi R}
\sum_n \psi_{k\,i}^{(n)}(x) f_{k\,i}^{(n)}(y)$
so that upon compactification on the orbifold $S^1/Z_2$ the zero 
modes, $\psi_{1\,i}^{(0)}$ will become the neutrino states $\nu_{L,R}$, 
while the zero modes, $\psi_{2\,i}^{(0)}$ are projected out.  At the massive 
Kaluza-Klein level each fermion $\psi_{1\,i}^{(n)}$ pairs up with the 
fermion $\psi_{2\,i}^{(n)}$ to form massive vector-like Dirac states. 
Assume that the action for the bulk fermion fields $\Psi_i$ in the 
warped space is given by
\begin{equation}
\label{totaction}
    S=-\int d^4 x \int dy \sqrt{-g} \left[ i \bar\Psi_i\Gamma^M D_M\Psi_i 
        + i c_i k\epsilon(y)\bar\Psi_i\Psi_i 
        + i \frac{b_M}{2} \delta(y)({\bar\Psi_R^c}{\Psi_R}+h.c.)\right]~,
\end{equation}
where $c_i,b_M$ are dimensionless constants which parametrise the bulk Dirac 
and boundary Majorana mass terms, respectively. 
The kinetic term in (\ref{totaction}) contains the gamma matrices
$\Gamma^M$ defined in curved space, and the covariant derivative $D_M
=\partial_M +\omega_M$, where $\omega_M$ is the spin connection (see 
Ref~\cite{gp1}).
When $b_M=0$ the canonically normalised zero mode wavefunction is 
determined to be~\cite{gp1}
\begin{equation}
         f_{1\,i}^{(0)}(y) = \frac{1}{N_0} e^{(2-c_i) k |y|}~,
\end{equation}
where $N_0$ is a normalisation constant. The specific value 
of $c_i$ parametrising the bulk Dirac mass term determines the degree of 
localisation of the zero mode field in the extra dimension. When $c_i>1/2\,
(c_i< 1/2)$ the zero mode is localised towards the Planck (TeV) brane.
The bulk Dirac mass term provides a convenient way to localise the zero mode
fields at any position in the extra dimension.

In this 5D geometrical framework suppose that the left-handed neutrino 
$\nu_L$ is localised towards the Planck brane and parametrised by the bulk 
mass parameter $c_L$, while the right-handed neutrino $\nu_R$ is localised 
towards the TeV brane, and parametrised by the bulk mass parameter 
$c_R$. On the Planck brane the source of lepton number 
violation will be parametrised by a Majorana mass term with nonzero $b_M$ in
(\ref{totaction}) for the right-handed neutrino. In the presence of this
lepton number violation the right-handed neutrino zero mode field will no 
longer remain massless. However, since the zero mode is assumed to be 
localised towards the TeV brane we expect the zero mode will only acquire a 
small mass depending on how well the field is localised.

To calculate the zero mode mass in the presence of the Majorana boundary term
we will consider the classical equations of motion for the right-handed 
neutrino Dirac spinor $\Psi_R$. These are given by
\begin{eqnarray}
\label{eom}
    &&i e^{k |y|} \bar\sigma^\mu\partial_\mu\widehat\nu_{2\,R} + 
       (\partial_5 + c_R k\epsilon(y)) \widehat{\bar\nu}_{1\,R} = 0~,\\
    &&i e^{k |y|} \bar\sigma^\mu\partial_\mu\widehat\nu_{1\,R} - 
       (\partial_5 - c_R k\epsilon(y)) \widehat{\bar\nu}_{2\,R} 
       + b_M \delta(y) \widehat{\bar\nu}_{1\,R} = 0~,
\end{eqnarray}
where we have absorbed the spin connection term by defining
$\widehat\nu_{iR} = e^{-2k |y|} \nu_{iR}$. Assuming that the solutions 
have the form $\widehat\nu_{iR}(x,y)=1/\sqrt{\pi R}\sum \nu_{iR}^{(n)}(x) \hat 
f_{iR}^{(n)}(y)$ we obtain~\cite{gp1}
\begin{eqnarray}
\label{solutions}
      f_{1\,R}^{(n)}(y) &=& \frac{e^{\frac{5}{2} k|y|}}{N_{1\,n}}\left[
       J_{\alpha_1}\left(\frac{m_n}{k e^{-k|y|}}\right) 
        -\frac{J_{\alpha_1+1}
       (\frac{m_n}{ke^{-\pi kR}})}{Y_{\alpha_1+1}(\frac{m_n}{k e^{-\pi kR}})}
       Y_{\alpha_1}\left(\frac{m_n}{ke^{-k|y|}}\right)\right],\\
      f_{2\,R}^{(n)}(y) &=& \epsilon(y) \frac{e^{\frac{5}{2}k|y|}}{N_{2\,n}}
       \left[J_{\alpha_2}\left(\frac{m_n}{ke^{-k|y|}}\right) 
       -\frac{J_{\alpha_2}(\frac{m_n}{ke^{-\pi kR}})}{Y_{\alpha_2}
        (\frac{m_n}{ke^{-\pi kR}})}
       Y_{\alpha_2}\left(\frac{m_n}{k e^{-k|y|}}\right)\right],
\end{eqnarray}
where $\alpha_{1,2}=|c_R\pm\frac{1}{2}|$, $N_{k\,n}$ are 
normalisation constants, and 
at $y=\pi R$ we have imposed even (odd) boundary conditions 
on the fermions, $f_{1\,R} (f_{2\,R})$. The Kaluza-Klein mass spectrum $m_n$ 
is obtained by taking into account the boundary Majorana mass at $y=0$,
and leads to the boundary condition
\begin{equation}
\label{planckbc}
    f_{2\,R}^{(n)}(0) = \frac{b_M}{2}~f_{1\,R}^{(n)}(0)~.
\end{equation}
This gives rise to the lightest mass mode
\begin{equation}
\label{zeromass}
    m_0 \simeq \frac{b_M}{2}~(1-2 c_R)k~e^{-(1-2c_R)\pi kR}~,
\end{equation}
where we are assuming $c_R < -1/2$. This formula is analogous to that
obtained for a boundary gaugino mass~\cite{mp,gp4}, and appears in
Refs~\cite{hs,csaki} which also considered a boundary Majorana neutrino mass. 
The masses of the Kaluza-Klein modes $\nu_{1\,R}^{(n)}$ and 
$\nu_{2\,R}^{(n)}$ which normally form Dirac pairs, can also be obtained by
solving (\ref{planckbc}). They are separated by a mass gap of order the 
TeV scale, and obtain an exponentially suppressed Majorana mass 
contribution from the boundary mass at $y=0$.

Thus, we see that in the presence of a boundary
Majorana mass the zero mode obtains a Majorana mass (\ref{zeromass}).
However, even though the lepton number violation is of order the
Planck scale, the Majorana mass (\ref{zeromass}) for the right 
handed neutrino is exceedingly small
assuming that $k e^{-\pi kR} =$ TeV. In fact if the right-handed neutrino is 
completely localised on the TeV brane, corresponding to the formal limit 
of $c_R\rightarrow -\infty$, then the right-handed neutrino remains massless.
This is because there are no direct couplings to the Planck brane, and 
any source of lepton number violation in the UV. Note that if instead the 
Majorana mass parameter on the Planck brane $b_M\rightarrow \infty$, then 
the lowest lying states 
are two degenerate singlet Weyl neutrinos forming a Dirac particle state 
with mass
\begin{equation}
\label{diracmass}
    m_0 \simeq \pm \sqrt{4 c_R^2 - 1}~k~e^{-(\frac{1}{2}-c_R)\pi kR}~,
\end{equation}
where $c_R< -1/2$. Even though this mass can be made arbitrarily small, 
it includes an extra Weyl neutrino $\nu_{2\,R}$ (besides $\nu_{1\,R}$).

We have seen that the right-handed Weyl neutrino $\nu_R$ remains massless
in the presence of lepton number violation on the UV brane. However 
we still need to generate a neutrino mass. This is accomplished by writing
a bulk Yukawa interaction between the left-handed neutrino $\nu_L$, 
right-handed neutrino $\nu_R$, and the Higgs field, $H$. This generates
a Dirac neutrino mass where we will assume that the right-handed neutrino 
is sufficiently localised on the TeV brane, so that its lepton number 
violating Majorana mass is negligibly small. 

For simplicity consider the original Randall-Sundrum 
model~\cite{rs}, where a completely localised Higgs field breaks electroweak 
symmetry on the TeV brane. This generates a Dirac mass term on the TeV brane 
from the Yukawa interaction
\begin{equation}
\label{yukint}
        \int d^4x \int dy~\lambda_5~\nu_L(x,y) \nu_R(x,y) 
      H(x)~\delta(y-\pi R)~,
\end{equation}
where $\lambda_5$ is the 5D Yukawa coupling.
When the Higgs field obtains a vacuum expectation value, 
$\langle H\rangle= v$, the effective four dimensional Yukawa coupling is
given by~\cite{gp1}
\begin{equation}
\label{4dyukawa}
      \lambda_4 = \frac{\lambda_5 k}{N_L N_R} e^{(1-c_L-c_R)\pi kR}
      \simeq ~\lambda_5 k \sqrt{(c_L-1/2)(1/2-c_R)}~e^{(\frac{1}{2}
      -c_L)\pi kR},
\end{equation}
where $1/N_i^2=(1/2-c_i)/(e^{(1-2c_i)\pi kR}-1)$, and
$c_L>1/2$, $c_R<-1/2$. 
Assuming that $\lambda_5 k \sim 1$, we see that the Dirac neutrino mass
\begin{equation}
       m_\nu = \lambda_4 v \sim 10^{-2} {\rm eV}~,
\end{equation}
is naturally obtained for $c_L \sim 1.36$, where we have taken $c_R <
-1/2$ and $\pi kR = 34.54$. Thus, in the IR where lepton 
number is a good symmetry the neutrino is a Dirac particle and the 
magnitude of its mass can naturally be obtained via a small wavefunction
overlap. Since the right-handed 
neutrino lives on the TeV brane
it is not sensitive to any UV violation of global lepton number symmetry.

Note also that the Yukawa coupling (\ref{4dyukawa}) 
does not exponentially depend on the right-handed bulk Dirac mass parameter 
$c_R$. This is because the right-handed field is assumed to be localised 
towards the TeV brane, so that the wave function overlap with the localised 
Higgs field is of order one. This of course means that in this simple setup
the Yukawa coupling will generically be the same for each fermion in the 
left-handed doublet unless there is a hierarchy in the bulk mass parameters
$c_{\nu_R}\sim 10^{-14} c_{e_R}$. However, a set up with no hierarchies can
easily be obtained by delocalising the Higgs field in a supersymmetric 
framework~\cite{gp1,gp2}. For example, requiring $c_H=0.5$, $c_L=1.32$,
$c_{e_R}\simeq 0.2$, and $c_{\nu_R}\simeq -2$, leads to an electron Yukawa
coupling, $\lambda_e\sim 10^{-6}$ and an electron-neutrino Yukawa coupling,
$\lambda_{\nu_e}\sim 10^{-13}$. Alternatively, for $c_H=0.11$, $c_L=1.32$ the 
Yukawa couplings are obtained for $c_{e_R}\simeq 1$ and $c_{\nu_R}\simeq 
-2.5$. A solution can always be found with $c_{\nu_R}$ large and negative,
so that the right handed neutrino is localised towards the TeV brane, away
from the UV violation of lepton number on the Planck brane.

\section{AdS/CFT correspondence}

Remarkably the 5D geometric picture can be given a purely 4D description
via the AdS/CFT correspondence~\cite{adscft}. This correspondence
relates the 5D AdS theory to a large $N$ strongly coupled 4D CFT.
The UV brane at $y=0$ corresponds to a UV cutoff at momentum $p=k$ in the 
4D CFT, while the boundary at $y=\pi R$ corresponds to an 
IR cutoff at $p=k e^{-\pi kR}$~\cite{apr,rzv}. In this way we see that 
the slice of AdS$_5$ provides a dual description to a slice in momentum 
space of the large $N$ strongly coupled CFT. The conformal symmetry
is spontaneously broken in the IR leading to the formation 
of CFT bound states, much like the mesons and baryons in QCD.
In fact, fields which are localised towards the TeV brane are 
interpreted as bound states of the CFT, while fields which are localised 
towards the Planck brane are understood to be fundamental states which must 
be added to the CFT. In particular the Kaluza-Klein spectrum which is always 
localised towards the IR brane, corresponds to the infinite number of 
bound states of the strongly coupled CFT which are weakly coupled for large 
$N$~\cite{witten2}. In general the $y=0$ boundary value of 5D bulk fields 
$\Phi$ are identified as sources of CFT operators $\cal O$ via the term
\begin{equation}
\label{sourceterm}
     {\cal L}= \lambda \Phi(x) {\cal O}(x)~,
\end{equation}
where the mass of the bulk field is related to the dimension
of the CFT operator. 

The case of global lepton number symmetry is very similar to what 
occurs with global supersymmetry~\cite{gp4}. In particular let us consider 
the case of a bulk fermion $\Psi$. 
Only half the degrees of freedom of the bulk Dirac fermion 
are identified as sources on the AdS boundary~\cite{spinors}. 
The four-dimensional theory 
consists of the CFT sector fields and the fundamental sector source fields. 
The physical mass eigenstates in the 4D dual theory are a linear
superposition of the CFT sector fields and the fundamental source fields.
The global lepton number symmetry is only a symmetry of the CFT sector,
whereas the fundamental sector explicitly breaks the global symmetry.

Consider first the bulk Dirac fermion field $\Psi_L$ 
containing the left-handed neutrino field. In the 4D dual theory we 
introduce the fundamental source field $\psi_L$ and for $c_L > 1/2$
we have
\begin{equation}
\label{lhdual}
     {\cal L}= {\cal L}_{CFT} + i\bar\psi_L {\bar\sigma}\cdot\partial\psi_L
         + \xi k^{1/2-c_L} \bar\psi_L{\cal O}_L + h.c. + \dots~,
\end{equation}
where $\xi$ is a dimensionless coupling and ${\cal O}_L$ is a composite
fermion operator which couples to $\psi_L$ with dim~${\cal O}_L=3/2
+|c_L+1/2|$~\cite{cnp}.
When $c_L>1/2$ the coupling which mixes the fundamental sector with the CFT 
sector is always irrelevant. From the 5D theory we know that the physical
neutrino $\nu_L$ is localised towards the Planck brane, so that in the dual 
theory this is interpreted by saying that the physical mass eigenstate is 
predominantly composed of the fundamental sector state, $\psi_L$.

Similarly for the bulk Dirac fermion field $\Psi_R$ containing the 
right-handed neutrino field. In this case the dual theory Lagrangian 
for $c_R<-1/2$ is given by
\begin{equation}
\label{rhdual}
     {\cal L}= {\cal L}_{CFT} + \xi k^{1/2+c_R} \bar\psi_R{\cal O}_R 
      + h.c. + \dots~,
\end{equation}
where $\psi_R$ is a fundamental source field which couples to the
CFT operator ${\cal O}_R$ with dim~${\cal O}_R= 3/2+|c_R-1/2|$~\cite{cnp}. 
The physical right-handed neutrino is localised towards the TeV brane, and 
we see that in the 4D dual theory the mixing with the fundamental source 
field $\psi_R$ is highly suppressed in the limit $c_R\rightarrow -\infty$.
This means that the physical mass 
eigenstate $\nu_R$ does not couple to the fundamental sector. Below the 
TeV scale the right-handed neutrino $\nu_R$ is a composite CFT bound state. 

Normally in four dimensions the dimension three operator (\ref{majterm}) 
will decouple the right-handed neutrino at low energies, and the light 
neutrinos are necessarily Majorana. Let us now understand why this is no 
longer the case in the 4D dual CFT theory. In the 5D theory lepton
number symmetry was only broken on the Planck brane, and using the
AdS/CFT dictionary, this corresponds to breaking lepton number symmetry at the 
Planck scale in the 4D dual description. Only the fundamental sector
can feel this explicit UV breaking, while the global lepton number symmetry is 
preserved in the CFT sector. This means that the composite right-handed 
neutrino will only feel the explicit breaking via the mixing term in 
(\ref{rhdual}), where for the fundamental state $\psi_R$ we include
a mass term 
\begin{equation}
    \frac{k}{2b_M}\psi_R\psi_R + h.c.
\end{equation}
In the CFT the operator 
${\cal O}_R$ has the appropriate quantum numbers to create the right-handed 
neutrino composite states, $\psi_{{\cal O}_R}^{(n)}$. We will assume that 
composite particle states will appear at the scale $\mu$, where the 
conformal symmetry is spontaneously broken. Thus by writing 
${\cal O}_R \simeq\mu^{1/2-c_R}~\psi_{{\cal O}_R}^{(0)}$  at
$q^2\simeq 0$ for the massless resonance, the mixing matrix with the 
fundamental field $\psi_R$ is given by
\begin{eqnarray}
  && (\bar\psi_R, \psi_{{\cal O}_R}^{(0)})
      \left(\begin{tabular}{cc}
             $\frac{k}{2b_M}$ &  $\frac{\xi}{2} k 
               (\frac{\mu}{k})^{\frac{1}{2}-c_R}$\\
             $\frac{\xi}{2} k (\frac{\mu}{k})^{\frac{1}{2}-c_R}$ & 0 
            \end{tabular}\right)
    \left(\begin{tabular}{c}
            $\bar\psi_R$\\ 
            $\psi_{{\cal O}_R}^{(0)}$
            \end{tabular}\right) + h.c.\nonumber\\
 &\simeq&  (\bar\psi^\prime_R, \psi_{{\cal O}_R}^{(0)\prime})
      \left(\begin{tabular}{cc}
             $\frac{k}{2b_M}+\dots$ &  0\\
             0 & $-\frac{\xi^2}{2} b_M~k(\frac{\mu}{k})^{1-2c_R}+\dots$
            \end{tabular}\right)
    \left(\begin{tabular}{c}
            $\bar\psi^\prime_R$\\ 
            $\psi_{{\cal O}_R}^{(0)\prime}$
          \end{tabular}\right) + h.c.~,
\end{eqnarray}
where we have assumed that $\mu\ll k$ and in terms of the original fields
we obtain
\begin{eqnarray}
       \bar\psi^\prime_R &\simeq& \bar\psi_R + b_M~\xi
        \left(\frac{\mu}{k}\right)^{\frac{1}{2}-c_R} 
        \psi_{{\cal O}_R}^{(0)}~,\\
       \psi_{{\cal O}_R}^{(0)\prime} &\simeq& \psi_{{\cal O}_R}^{(0)} 
       - b_M~\xi \left(\frac{\mu}{k}\right)^{\frac{1}{2}-c_R} \bar\psi_R~.
\end{eqnarray}
Thus, the dimension three 
Majorana mass operator at the scale $\mu$ becomes for $c_R< -1/2$
\begin{equation}
      {\cal L}_{Majorana} \simeq b_M \left(\frac{\mu}{k}
        \right)^{1-2c_R}~k~\nu_R\nu_R+h.c.~,
\end{equation}
where $\nu_R$ is associated with the composite field
$\psi_{{\cal O}_R}^{(0)\prime}$. 
Assuming that $\mu=k e^{-\pi kR}$ then we obtain the result consistent with 
(\ref{zeromass}). In the limit that $c_R\rightarrow -\infty$ we see that the 
physical right-handed neutrino is mostly a  composite state, and 
the Majorana mass is highly suppressed so that the dimension 
three operator becomes irrelevant in the IR. Consequently, at low energies 
global lepton number symmetry is restored.

In the case where the boundary Majorana mass parameter 
$b_M\rightarrow\infty$, the zero mode does not couple to the Planck brane
which means that the fundamental source field is massless. The mixing matrix 
becomes
\begin{equation}
   (\bar\psi_R, \psi_{{\cal O}_R}^{(0)})
      \left(\begin{tabular}{cc}
             $0$ &  $\frac{\xi}{2} k(\frac{\mu}{k})^{\frac{1}{2}-c_R}$\\
             $\frac{\xi}{2} k(\frac{\mu}{k})^{\frac{1}{2}-c_R}$ & 0 
            \end{tabular}\right)
    \left(\begin{tabular}{c}
            $\bar\psi_R$\\ 
            $\psi_{{\cal O}_R}^{(0)}$
            \end{tabular}\right) + h.c.~,
\end{equation}
and give rises to masses $\pm\frac{\xi}{2}~k~e^{-(\frac{1}{2}-c_R)
\pi kR}$, which form a Dirac mass that is consistent with the bulk 
calculation (\ref{diracmass}).

Finally, using the AdS/CFT dictionary it is straightforward to understand the 
smallness of the Dirac neutrino Yukawa coupling in the 4D dual theory. 
The Higgs is assumed to be localised on the TeV brane and corresponds
to a CFT bound state much like the right-handed neutrino.
In a large $N$ gauge theory the trilinear meson vertex $\Gamma_3$,
corresponding to a Yukawa interaction between three composite states 
is $\Gamma_3\sim1/\sqrt{N}$~\cite{witten2}. However, the left-handed 
neutrino state is localised towards the Planck brane, which means that 
the physical left-handed neutrino is primarily composed
of the fundamental state. Thus, in the Yukawa interaction we need
to take into account the mixing term (\ref{lhdual}) and the
matrix element $\langle 0|{\cal O}_L|\psi_{{\cal O}_L}^{(n)}\rangle 
\sim \sqrt{N}$. At the scale $\mu$ this leads to the Yukawa interaction
\begin{equation}
   {\cal L}_{Yukawa}\simeq\left(\frac{\mu}{k}\right)^{c_L-\frac{1}{2}} 
    \nu_L\nu_R H +h.c.
\end{equation}
Assuming that the conformal symmetry is broken at the scale $\mu=k 
e^{-\pi kR}$ then we obtain the four-dimensional Yukawa coupling consistent
with the bulk calculation (\ref{4dyukawa}). In the dual CFT picture we see 
that the Yukawa coupling is naturally suppressed because the mixing between 
the fundamental sector and CFT sector is very small. It is also easy to see
that the suppression only depends on $c_L$ or the localisation of the 
left-handed
neutrino. The CFT operators ${\cal O}_i$ create the composite bound states
and since both the Higgs and right-handed neutrino are primarily CFT bound 
states the mixing contribution from the fundamental sector is extremely 
suppressed for these fields. Again this is similar to what happens for 
global supersymmetry~\cite{gp4}.

\section{Conclusion}

If the standard model is augmented with a right-handed singlet
neutrino ($\nu_R$) then we have shown that the neutrino masses can be 
purely Dirac even in the presence of an explicit UV breaking of lepton 
number. In five dimensions this can be understood by requiring that 
lepton number is explicitly broken on the UV brane while the right-handed 
neutrino is localised on the IR brane. The fact that the right-handed
neutrino is physically separated in the bulk from the source of lepton
number violation means that the lepton number symmetry is preserved at low
energies. Remarkably there exists a purely four-dimensional description
of this model. At the TeV scale the right-handed neutrino is predominantly
a CFT bound state which contains a tiny admixture of a fundamental state. 
In the CFT sector the lepton number symmetry is preserved while the symmetry 
is explicitly broken in the fundamental sector. The extremely tiny mixing
with the fundamental sector accounts for the preservation of lepton number
symmetry at low energies. In this setup there is no need to introduce an
intermediate mass scale since the neutrino masses are purely Dirac, and
the smallness of the Yukawa couplings is explained by a small wavefunction
overlap or large anomalous dimensions of CFT operators. In addition 
our model differs from previous Dirac neutrino models in extra dimensions 
in that we do not assume that the fundamental theory in the UV preserve 
a global lepton number symmetry. This removes the theoretical impediment 
of requiring that a consistent theory of quantum gravity preserve a global
quantum number, and provides an alternative to the usual seesaw mechanism 
of Majorana neutrino masses.

\section*{Acknowledgments}
I would like to thank K.S. Babu, A. Pomarol, M. Shifman, A. Vainshtein and
M. Voloshin for useful discussions. This research was supported in part 
by a DOE grant DE-FG02-94ER40823 at the University of Minnesota, and by 
a grant from the Office of the Dean of the Graduate School of the 
University of Minnesota.

\end{document}